\documentclass[aps,prb,twocolumn,showpacs]{revtex4}
\usepackage{tabularx,graphicx}
\begin{document}
%\draft
%\twocolumn[\hsize\textwidth\columnwidth\hsize\csname@twocolumnfalse%
%\endcsname
\title{Interlayer hopping properties of electrons in layered metals}
\author{M.A.H. Vozmediano$^{\dag}$,
M. P. L\'opez-Sancho$^{\ddag}$ and F. Guinea$^{\ddag}$.}
\affiliation{
$^{\dag}$ Departamento de Matem\'aticas,
Universidad Carlos III de Madrid, Avda. de la Universidad 30,28911 Legan\'es, 
Madrid, Spain. \\                                                             
$^{\ddag}${I}nstituto de Ciencia de Materiales de Madrid,
CSIC, Cantoblanco, E-28049 Madrid, Spain.}
\date{\today}
%\maketitle
\begin{abstract}
A formalism is proposed to study the electron tunneling between extended
states, based on the spin-boson Hamiltonian previously used in
two-level systems. It is applied to analyze the out--of--plane tunneling in 
layered metals considering
different models. By studying the effects  of in--plane interactions 
on the interlayer tunneling of electrons near the Fermi level,
we establish the relation between departure from Fermi liquid behavior 
driven by electron correlations inside the layer and the out of plane 
coherence. Response functions, directly comparable with experimental 
data are obtained.

\end{abstract}
\maketitle
\section{Introduction}
%{\it Introduction}.
Layered materials have been object of intensive study
since they present important physics. 
A key feature of many layered materials is the anisotropy exhibited
by its transport properties: while being metallic within the layers, 
the transport in the $c$-axis, perpendicular to the layers, may be 
coherent or incoherent and undergo a crossover with temperature from
one regime to the other, thus changing the effective 
dimensionality of the system\cite{valla, esq0}.
Correlation effects increase as dimensionality decreases, therefore 
dimensionality is crucial for the electronic properties and to choose
the appropriate model to study the system.
Unusual properties
are derived from the anisotropy and periodicity along the
axis perpendicular to the planes i.e. the structure of
collective excitations absent in two dimensional (2D) and
three dimensional (3D) electron gases \cite {quinn}.

Among the most studied layered materials are the  
high-temperature cuprate superconductors.
These compounds present a strong anisotropy 
and are treated as two-dimensional systems in many approaches.
In the normal state the transport properties within the doped CuO$_2$
planes are very different from those along the $c$-axis: electron
motion in the $c$-direction is incoherent in contrast with the 
metallic behavior of the in-plane electrons as probed by the 
different {$\bf \rho_{c}$} and {$\bf \rho_{ab}$} resistivities
and their different dependence with temperature\cite {fuji,ando}.
Optical conductivity measurements confirm the anomalous c-axis 
properties \cite {review}. 
These features have led to the 
asumption that the relevant physics of these substances lies on the 
CuO$_2$ planes, common to all the families. However, the importance
of the c-axis structure is  being intensively investigated 
since it offers a key to explain the differences of the critical
temperature T$_c$ and optimal doping in the different compounds. 
The relevance of the nature of the conductance in the
direction perpendicular to the CuO$_2$ planes to the nature
of the superconducting phase has been remarked
on  both theoretical \cite {ander,legg,TL01}
and  experimental \cite {marel} grounds.
The anomalous behavior of the out of plane properties in the cuprates,
and in analogy with the one dimensional Luttinger liquid, 
has led to the proposal of the
failure of the conventional Fermi-liquid (FL) theory in these compounds\cite{ander}.

Recent $c$-axis transport measurement focused on 
the resistivity anisotropy {$\bf \rho_{c}$}/{$\bf \rho_{ab}$} ratio indicate
that {$\bf \rho_{c}$}/{$\bf \rho_{ab}$} quickly increases with decreasing 
temperature. It was remarkably found that, at moderate temperature 
(between 100 to 300K), {$\bf \rho_{c}$}/{$\bf \rho_{ab}$} is almost 
completely independent of doping $x$  in the non superconducting regime
($0.01\le x \le 0.05$), which suggest that the same charge confinement
mechanism that renormalizes the $c$-axis hopping rate survives down to
$x=0.01$. On the contrary the {$\bf \rho_{c}$}/{$\bf \rho_{ab}$} ratio
quickly changes when x enters into the superconducting regime, indicating
that interlayer hopping quickly 
becomes easier as $x$ is increased above 0.05, i.e. the charges
are increasingly less confined in this doping range\cite{Komiya}.
Measurements of the out-of-plane magnetoresistance(MR) of 
La$_{2-x}$SR$_x$ CuO$_4$
over a wide doping range and comparison with other systems
suggest that a competition between the $c$-axis hopping rate and
the in-plane scattering rate determines the behavior of the $c$-axis MR
in layered perovskites with incoherent $c$-axis resistivity\cite{Huss}.   
             
Many alternative models have been    
suggested to explain the puzzling properties, from  Fermi liquid modified 
by strong electron-electron correlations \cite {cooper} to the 
non-Fermi Luttinger liquid \cite {ander}.
An alternative explanation of the 
emergence of incoherent behavior in the out of plane direction has been
proposed in terms of the coupling of the interlayer electronic motion to
charge excitations of the system\cite{TL01,VLG02}. This approach implicitly 
assumes that electron-electron interactions modify the in-plane
electron propagators in a non trivial way, at least at distances
shorter than the elastic mean free path. The strong angular dependence
of both the scattering rate and the interplane tunneling element
can also lead to the observed anisotropies\cite{IM98}.

Graphite, another layered material, presents an intraplane
hopping much larger than the interplane hybridization. 
Many of the transport properties established in the past for
this well known material are being questioned at present.
Recent conductivity measurements reveal a suppresion of 
the $c$-axis conductivity much larger than what would be
predicted by the band calculations of the interlayer hopping\cite{esqui1}.
Band structure calculations are also challenged by very
recent claims of observation of quantum Hall plateaus in
pure graphite\cite{esqui2}.
The unconventional transport properties of graphite such
as the linear increase with energy of the inverse 
lifetime\cite{Yetal96} (see also\cite{STK01}),
suggest deviations from the conventional Fermi liquid 
behavior, which could be due to strong Coulomb
interactions unscreened because of the lack of states at the
Fermi level \cite{GGV94,GGV99}.

By assuming that electron correlations modify the in-plane
electron propagators\cite{TL01}, we show that even 
in the clean limit, many-body effects 
can suppress the coherent contribution to the out of plane electron
hopping. The clean limit is defined  as that in which the 
elastic mean free path diverges.
It is shown that for certain 
models of correlated electrons\cite{GGV94,GGV99,GGV96} the
interplane hopping between extended states can be a relevant
or an irrelevant variable, in the Renormalization Group (RG) sense, 
depending on the strength of the coupling constant.       
The scheme used here is based on the RG analysis
as applied to models of interacting electrons\cite{S94,P92}. 
We will demonstrate the usefulness of the method, by rederiving
known results for a non trivial system, 
the array of Luttinger liquids\cite{W90,KF92,1D}, and then we will
apply the method to two well defined two-dimensional systems where
ordinary perturbation theory fails, due to the existence of logaritmic
divergences. 

Recently there has been renewed interest in two--dimensional
systems that support low--energy excitations which can be described
by Dirac fermions. Examples are the flux--phase of planar 
magnets\cite{KL99,RW01}, nodal quasiparticles in d--wave 
superconductors\cite{FT01,Y01}, the insulating spin density wave
phase of high T$_c$ superconductors \cite{H02}, and quasiparticles of 
planar zero--gap
semiconductors as graphite\cite{GGV94,GGV99}. Long range Coulomb interactions
and disorder are modelled in these systems by means of gauge fields 
coupled to the Dirac fermions. Most of these systems show anomalous
transport properties ranging from mild departures of Fermi liquid 
behavior as in pure graphite, to the total destruction of the
quasiparticle pole. Disorder can be modelled as random gauge
fields coupled  to the Dirac quasiparticles. It affects the 
quasiparticle Green's function in a computable way what in turn 
modifies the interlayer tunneling.

%For the problem of interchain hopping 
%between Luttinger liquids, it recovers well known 
%results\cite{W90,KF92,1D}.
This paper is organized as follows. 
In section II we present the method of calculation and show the results it
gives in one dimension. In section III we apply it to 
two-dimensional models which show deviations from FL behavior: in particular
to systems with a vanishing density of states at the Fermi level 
as is the case of graphene sheets and to a system of planar electrons near
a Van Hove singularity. The effect of disorder on these graphitic
and related systems is discussed.
The main physical consequences of our calculation are presented
in section IV.
%In the present work we propose a microscopic model which provides 
%the metallic nature of the CuO$_2$ planes and the incoherent transport in
%the out of plane direction including only repulsive electron-electron
%interactions \cite {gonz}.

\section{The method of calculation.}
%{\it The method of calculation.}
In the presence of electron-electron interactions, tunneling processes
are modified by inelastic scattering events.
The influence of inelastic scattering on electron tunneling
has been studied, using equivalent methods, in mesoscopic devices
which show Coulomb blockade\cite{SET}, Luttinger
liquids\cite{W90,KF92,1D,SNW95}, and dirty metals\cite{RG01}.
The simplest formulation of the method replaces the
excitations of the system (such as electron-hole      
pairs) by a bath of harmonic oscillators
with the same excitation spectrum.
This approach can be justified rigorously in one dimension,
and is always an accurate description of the response of
the system when the coupling of the quasiparticles to each individual
excitation is weak, although the net effect of the environment
on the system under study can be large\cite{CL83}. 
The expression for the coupling
between the electrons and the oscillators is obtained by assuming that the oscillators
describe the charge oscillations of the system. Then, the coupling can be related,
using perturbation theory, to the charge-charge correlations in the electron gas. This
is consistent with the assumption that the modes in the environment are weakly perturbed 
by their interaction to the low energy electrons whose tunneling properties are been considered: 
                                                                           
\begin{equation}
{\cal H}_{int} = c^\dag_i c_i \sum_{\bf \vec{k}}
V_i ( {\bf \vec{k}} ) \hat{\rho}_{\bf \vec{k}}
\label{int}
\end{equation}
where $c^\dag_i$($c_i$)  creates(destroys) an electron at site $i$,
and $\hat{\rho}_{\bf \vec{k}}$ describes the charge
fluctuations of the environment, which are to be described
as a set of harmonic modes as stated above.
The interaction in Eq.(\ref{int}) is the simplest, and most common,
coupling between the tunneling electrons and the excitations of the system,
which is spin independent.
Other spin-dependent or more complicated couplings can also
be taken into account, provided that the appropriate
response function is used.                                                                       
            
\begin{figure}
\resizebox{6cm}{!}{\includegraphics[width=7cm]{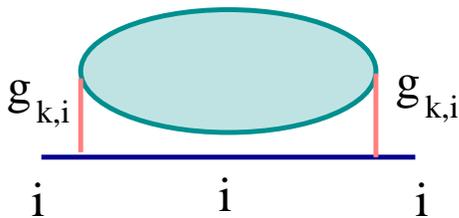}} 
\caption{Diagram corresponding to the lowest order contribution to
the self energy of an electron at site $i$ in the hamiltonian given
in Eq.(\protect\ref{hamil}). This diagram can be used to describe
the spectral function in Eq.(\protect\ref{J_w}).}
\label{diagram}
\end{figure}

Since the excitations of the 
system (electron-hole pairs, plasmons) are modelled as bosonic
modes, one can write an effective electron-boson hamiltonian of the type:
\begin{eqnarray}
{\cal H}_{e-b} &=&{\cal H}_{elec} + {\cal H}_{env} + {\cal H}_{int} \nonumber
 \\                                                                    
&= &\sum {\bf t}_{ij} c^\dag_i c_j + \sum \omega_k b^\dag_k
b_k + \sum g_{k , i} c^\dag_i c_i ( b^\dag_k + b_k )  
\label{hamil}
\end{eqnarray}
where ${\cal H}_{elec}$ describes the individual quasiparticles,
${\cal H}_{env}$ stands for the set of harmonic oscillators
which describe the environment, and ${\cal H}_{int}$ defines
the (linear) coupling between the two.
The $b_k^\dag$($b_k$)
are boson creation(destruction)  operators,
the ${\bf t}_{ij}$ describe the electronic hopping processes. 
The
information about the interaction between the electron in
state $i$ and the
environment is encoded in the  spectral function\cite{CL83} 
\begin{equation}
J_i ( \omega ) = \sum_k
| g_{k,i} |^2 \delta ( \omega - \omega_k )
\label{J_w}
\end{equation}
The function $J_i ( \omega )$ describes the retarded interaction induced by the
environment. In standard perturbative treatments of the hamiltonian
in Eq. (\ref{hamil}), this function is simply the self energy of an electron at site
$i$, as schematically shown in Fig.[\ref{diagram}].

Using second order perturbation theory and Eq.(\ref{int}),
we can write\cite{CL83,SET}:
\begin{equation}
J_i ( \omega ) = \sum_{\bf \vec{k}} V_i^2 ( {\bf \vec{k}} )
{\rm Im} \chi ( {\bf \vec{k}} , \omega )
\end{equation}
where $\chi ( {\bf \vec{k}} , \omega )$ is the Fourier transform
of the density-density response of the system, 
$\langle \hat{\rho}_{\bf \vec{k}} ( t ) \hat{\rho}_{\bf - \vec{k}}
( 0 ) \rangle$. 
The electron-boson interaction leads to a Franck-Condon factor which
reduces the effective tunneling rate. The Franck-Condon factor depends
exponentially on the coupling between the particles and the oscillators, and
when it diverges additional self consistency requirements between the hopping
amplitudes and the oscillator frequencies included in the calculation have 
to be imposed.
The electron propagators
acquire an anomalous time, or energy, dependence, that can be calculated to
all orders if the state $i$ is localized, or, which is equivalent, neglecting
the hopping terms in Eq.({\ref{hamil}). The present scheme can be considered 
a generalization to layered systems
of the approach presented in reference\cite{KF92} for a Luttinger liquid,
where the scaling of tunneling is expressed in terms of an effective density
of states. The tunneling will become irrelevant, as in \cite{KF92}, 
in one of the models  
considered here, that of electrons in the vicinity of a saddle
point of the dispersion relation, which is consistent with the picture
of the $c$-axis interlayer tunneling strongly suppressed by voltage 
fluctuations, proposed by Turlakov and Leggett \cite{TL01}. 

Using the boson representation of the environment, we can write 
the electron propagator as: 
\begin{eqnarray}
\langle c^\dag_i ( t ) c_i ( t' )
\rangle
&\sim  &\langle c^\dag_i ( t ) c_i ( t' ) \rangle_0 \times
\nonumber \\ & &\exp \left\{ -
\int d \omega  \left[ 1 - e^{i \omega ( t - t' )} \right]
\frac{J_{i} ( \omega )}
{ \omega^2} \right\}
\label{Green}
\end{eqnarray}
where $\langle c^\dag_i ( t ) c_i ( t' ) \rangle_0 \sim
e^{i \varepsilon_i ( t - t')}$ is the Green's function in
the absence of the interaction. This result can be viewed as the
exponentiation of the second order self energy diagram shown
in Fig.[\ref{diagram}]. The approximation of the excitations
in the environment by bosonic modes allows us to sum 
an infinite set of diagrams like the one in Fig.[\ref{diagram}].

In contrast to previous work, we  analyze tunneling between coherent
extended states. In order to do so, we need to
generalize Eq.(\ref{Green}) to this case. We then need
to  know the Green's function
of coherent states in the individual layers, $G( {\bf \vec{k}}, w)$, including
the correction due to the interaction to the environment. 
We firstly  assume
that Eq.(\ref{Green}) also holds in a system
with extended states. For a standard metallic system,
we must insert $\langle c^\dag_i ( t ) c_i ( t' ) \rangle_0 \sim
1 / ( t - t' )$ in Eq.(\ref{Green}). It can be shown that
this approximation is exact at short times,
$W \ll ( t - t' )^{-1} \ll \Lambda$, where $W$ is an energy scale
related to the  dynamics of the electrons, and $\Lambda$ is
the upper cutoff in the spectrum of the environment.            
%\begin{eqnarray}
%\langle c^\dag_i ( t ) c_j ( t' )
%\rangle &
%\sim &\frac{C_{ij}}{t - t'} \times \nonumber \\ &\times &\exp \left\{
%\int_{{\bf \vec{r}} ,
% {\bf \vec{r}'}}
% \int_{ \Omega}
%d^2 {\bf \vec{r}} d^2 {\bf \vec{r}'}
%d^2 {\bf \vec{k}} e^{i {\bf \vec{k}}
%( {\bf \vec{r} - \vec{r}'} )} \right. \nonumber \\ & & \left.
%\int d \omega \left[ 1 - e^{i \omega ( t - t' )} \right]
%\frac{V_{eff} ( {\bf \vec{k}} , \omega )}
%\frac{J_{ij} ( \omega )}
%{ \omega^2} \right\}
%\label{Green}
%\end{eqnarray}
%where:
%\begin{equation}
%J_{ij} (  \omega ) = \sum_k | g_{i k} g_{j k} |^2
%\delta ( \omega - \omega_k )
%\end{equation}
This expression can be generalized, taking into account the
spatial structure of the coupling to:
\begin{eqnarray}
\langle \Psi^{\dag}_i ( t ) \Psi_j ( t' )
\rangle &
\sim &\frac{1}{t - t'} \times \nonumber \\ &\times &\exp \left\{
 \int_{{\bf \vec{k}}}
 \int_{ \Omega} \int_{ \Omega }
d^2 {\bf \vec{r}} d^2 {\bf \vec{r}'}
d^2 {\bf \vec{k}} e^{i {\bf \vec{k}}
( {\bf \vec{r} - \vec{r}'} )} \right. \nonumber \\ & & \left.
\int d \omega \left[ 1 - e^{i \omega ( t - t' )} \right]
\frac{V_{eff} ( {\bf \vec{k}} , \omega )}
{ \omega^2} \right\}
\label{propagator}
\end{eqnarray}
where $\Omega$ is the region of overlap of the wavefunctions
$\Psi_i ( {\bf \vec{r}} )$ and 
$\Psi_j ( {\bf \vec{r}} )$.
This expression, which can be seen as the exponential of the leading
frequency dependent self--energy correction to the electron propagator,
has been extensively used in studies of tunneling in zero dimensional 
systems (single electron transistors) which show Coulomb 
blockade\cite{SET}, one dimensional conductors\cite{SNW95}, and
disordered systems in arbitrary dimensions\cite{RG01}.

The effective interaction can, in turn, be written in terms
of the response function as:
\begin{equation}
V_{eff} ({\vec {\bf k}} , \omega ) = V^2 ( {\vec {\bf k}} )
{\rm Im} \chi ({\vec {\bf k}} , \omega )
\label{veff}
\end{equation}

The time dependence in Eq. (\ref{propagator}) is determined
by 
%$\lim_{\omega \rightarrow 0} \chi ( {\bf \vec{k}} , \omega )$.     
the low energy limit of the response function.
In a Fermi liquid, we have
\begin{equation}
{\rm Im} \chi {\vec ({\bf k}} , \omega ) \approx \alpha ( {\vec{\bf k}} )
\vert \omega \vert \;\;\;  \omega \ll E_F\      \;,
\end{equation}
where $E_F$ is the Fermi energy.
%\end{equation}
Eq. (\ref{propagator}) then gives:
\begin{equation}
\lim_{(t-t') \rightarrow \infty} \langle \Psi^{\dag} ( t ) \Psi ( t' )
\rangle 
\sim \frac{1}{( t - t' )^{1 + {\alpha}}} 
\label{green}
\end{equation}
where
\begin{equation} 
{\alpha} = \int_{{\bf \vec{k}}}\int_{\Omega} \int_{\Omega} 
d^2 {\bf \vec{r}} d^2 {\bf \vec{r}}' 
d^2 {\bf \vec{k}} e^{i {\bf \vec{k}}
( {\bf \vec{r} - \vec{r}'} )} V^2( {\bf \vec{k}} ) \alpha ( {\bf \vec{k}} )
\label{alpha}
\end{equation}
The parameter ${\alpha}$ gives the correction to the
scaling properties of the Green's functions. Integration in ${\bf \vec{k}}$
is restricted to $ | {\bf \vec{k}} | \ll L^{-1}$
where $L$ is the scale of the region where the tunneling
process takes place. The value of $L$ is limited  by the
length over which the phase of the electronic wavefunctions
within the layers is well defined.

The former approach is directly related to the theoretical
model proposed in \cite{TL01} where the limit of coherent
tunneling is set by the value $\alpha=1$. In sec. IIIB
this procedure is applyed to the study of interlayer tunneling
in layered systems whose Fermi surface is close
to a saddle point which has been proposed as a possible
explanation of some properties of the cuprates.

We can also use Eq.(\ref{green}) to analyze the
interlayer tunneling by applying Renormalization Group methods.
The simplest case where this procedure has been used is for
the problem
of an electron tunneling between two
states, $i$ and $j$, which has been intensively
studied\cite{Letal87,W93}.  We integrate out
the high energy bosons, with energies
$\Lambda - d \Lambda \le
\omega_k \le \Lambda$ and rescaled
hopping terms are defined. As mentioned earlier,
Eq.(\ref{green}) is valid for this range of energies.
The renormalization of
the hoppings is such that the properties of the
effective Hamiltonian at energies $\omega \ll \Lambda$
remain invariant. If the hoppings ${\bf t}_{ij}$ are small, any
physical quantity which depends on them can be       
expanded, using time dependent perturbation theory, in powers of:
\begin{equation}
{\bf t}_{ij}^2 \langle c_i^\dag ( t ) c_j ( t ) c_j^\dag ( t' ) c_i ( t' )
\rangle \approx {\bf t}_{ij}^2 \langle c_i^\dag ( t ) c_i ( t' ) \rangle
\langle c_j ( t ) c_j^\dag ( t' ) \rangle
\label{perturbation}
\end{equation}
The integration of the high energy modes implies that the terms
in Eq.(\ref{perturbation}) are restricted to $t \le \Lambda^{-1}$,
or, alternatively, the time  unit have to be rescaled\cite{C81},
$\tau' = \tau e^{d\Lambda / \Lambda }$, where $\tau \sim
\Lambda^{-1}$.
Using Eq.(\ref{green}), the condition of keeping the
perturbation expansion in powers of the terms in Eq.(\ref{perturbation})
invariant implies that:
\begin{equation}
{\bf t}_{ij}^2 \rightarrow {\bf t}_{ij}^2  e^{
\frac{d \Lambda}{\Lambda} \left( 2 + 2 \alpha \right)}
\end{equation}
which can also be used to define the scaling dimension of
the hopping terms. The beta function of the hopping is then
\begin{equation}
\frac{\partial ( {\bf t}_{ij} / \Lambda )}{\partial l} \equiv
\beta( {\bf t}_{ij})=- \alpha \frac{ {\bf t}_{ij} }{\Lambda}
\label{renor}
\end{equation}
where $l = \log ( \Lambda_0 / \Lambda )$, and $\Lambda_0$
is the initial value of the cutoff.                                
  
This approach has been successfully used to describe inelastic
tunneling in different situations in\cite{TL01,W90,KF92,1D,SET,SNW95,RG01}.
 
The analysis which leads to Eq.(\ref{renor}) can be generalized to
study hopping between extended states, provided that we
can estimate the long time behavior of the Green's
function, 
\begin{equation}
G ( {\bf \vec{k}} , t - t' )
= \langle c^\dag_{\bf \vec{k}} ( t ) c_{\bf \vec{k}} ( t' )
\rangle .
\label{gtk}
\end{equation}

We assume that, in a translationally invariant system,
there is no dependence on the position of the local orbital, $i$.
This result implies that the frequency dependence of
the Green's function, in a
continuum description, can be written as:
\begin{equation}
\lim_{|{\bf \vec{r}} - {\bf \vec{r}}' | \rightarrow 0}
G ( {\bf \vec{r}} - {\bf \vec{r}}', \omega )
\propto | \omega |^{ \alpha}
\label{green_w}
\end{equation}
Equation(\ref{gtk}) is related to  Eq.(\ref{green_w}), by:
\begin{equation}
\lim_{|{\bf \vec{r}} - {\bf \vec{r}}' | \rightarrow 0}
G ( {\bf \vec{r}} - {\bf \vec{r}}', \omega ) =
\int d^D {\bf \vec{k}} G ( {\bf \vec{k}} , \omega )
\label{integral}
\end{equation}
where $D$ is the spatial dimension.
In the cases discussed below, the interaction is instantaneous in time,
and the non interacting Green's function can be written as:
\begin{equation}
G_0 ( {\bf \vec{k}} , \omega ) \propto \frac{1}{\omega}
{\cal F} \left( \frac{k_i^z}{\omega} \right)
\label{scaling_g0}
\end{equation}
where $z=1,2$ depending on the dispersion relation of the model.
In the following, we
assume that the interacting Green's function has the
same scaling properties, with the factor $\omega^{-1}$
replaced by $\omega^{-\delta}$ in Eq.(\ref{scaling_g0}), where $\delta$
depends on the interactions.
This can be shown to be correct
in perturbation theory to all orders,in the models                      
studied below, 
%which describe the physics in the proximity
%of critical points, 
because the self energy diagrams have a logarithmic dependence 
on $\omega$ (it is a well known fact for the
Luttinger liquid). Then, using eqs.
(\ref{green_w}), (\ref{integral})
and (\ref{scaling_g0}), we obtain:
\begin{equation}
G ( {\bf \vec{k}} , \omega ) \propto
| \omega |^{\alpha - D/z} {\cal F} \left( \frac{k_i^z}{\omega} \right)
\label{green_k}
\end{equation}
and $ {\cal F} ( u )$ is finite.
Thus, from the knowledge of
the real space Green's function, using
Eq.(\ref{Green}),  we obtain $\alpha$, which, in turn, determines
the exponent $\alpha - D / z $ which characterizes
$G ( {\bf \vec{k}} , \omega )$. Generically, we can write:
\begin{equation}
G_{loc,ext} ( \omega ) \sim | \omega |^{\delta_{loc,ext}}
\label{scaling_green}
\end{equation}
where the subindices $loc , ext$ stand for localized and extended
wavefunctions. In terms of these exponents, we can generalize
Eq.(\ref{renor}) to tunneling between general states to:
\begin{equation}
\frac{\partial ( {\bf t}_{ij}^{loc,ext} / \Lambda )}{\partial l} =
- \delta_{loc,ext} \frac{ {\bf t}_{ij} }{\Lambda}
\label{renor2}
\end{equation}                                                         

%Note that the scaling
%invariance of the Green's function, Eq.(\ref{scaling_g0}),
%is not a generic feature of Fermi liquids in dimensions
%greater than one.
Before proceeding to calculations of $\delta_{loc}$
and $\delta_{ext}$ for various models,
it is interesting to note that, in general,
the response function of an electron gas in dimension $D > 1$ behaves as
$$\lim_{\omega \rightarrow 0 , | {\bf \vec{k}} | \rightarrow 0}
\chi( {\bf \vec{k}} , \omega )
 \sim | \omega | / | {\bf \vec{k}} | ,$$ 
so that, from
Eq.(\ref{alpha}), 
$$\lim_{L \rightarrow
\infty} \alpha \sim L^{(1-D)}.$$
Thus, for $D > 1$, the contribution of the inelastic processes
to the renormalization of the tunneling vanishes for delocalized
states, $L \rightarrow \infty$. This result is consistent with the existence
of Fermi liquid behavior above one dimension\cite{MCC98}. Note that 
anisotropic Fermi surfaces, with inflection points, can lead
to marginal behavior above one dimension\cite{GGV97,FG02}.                             

It is easy to show that, in an isotropic Fermi liquid in D dimensions,
$\lim_{L \rightarrow \infty}
{\alpha} \propto L^{1 - D}$, where $L$ is the linear
dimension of the (localized) electronic wavefunctions $\Psi
( {\bf \vec{r}} )$. This result is due to
the dependence on ${\bf \vec{k}}$ of the response
functions which goes as 
$${\rm Im} \chi ( {\bf \vec{k}} ,
\omega ) \sim\frac{ | \omega | }{  k_F^{D-1} | {\bf \vec{k}} | }.$$
Thus, for $D > 1$, we recover coherent tunneling
in the limit of delocalized wavefunctions.

In one dimension, one can use the non interacting 
expression for ${\rm Im} \chi_0 ( {\bf \vec{k}} ,
\omega )$, to obtain:
\begin{eqnarray} 
& &\int
d^2 {\bf \vec{r}} d^2 {\bf \vec{r}'}
d^2 {\bf \vec{k}} e^{i {\bf \vec{k}}
( {\bf \vec{r} - \vec{r}'} )}
\int d \omega \left[ 1 - e^{i \omega ( t - t' )} \right]
\frac{V_{eff} ( {\bf \vec{k}} , \omega )}
{ \omega^2} \nonumber \\ 
&\propto &\left( \frac{U}{E_F} \right)^2 \times \left\{
\begin{array}{lr} 0 &t - t' \ll L / v_F \\
\log [ v_F ( t - t' ) / L ] &t - t' \gg L / v_F \end{array}
\right.
\label{1D}
\end{eqnarray}
where we have assumed a smooth short range interaction, parametrized by $U$.
Hence, the Green's functions have a
non trivial power dependence on time, even in the
$L \rightarrow \infty$ limit, in agreement with 
well known results for Luttinger liquids\cite{SNW95}.
In order to obtain the energy dependence of the effective
tunneling between ${\bf \vec{k}}$ states near the Fermi surface, one
needs to perform an additional integration over $d {\bf \vec{r}}$.
In general, near a scale invariant fixed point,
$\omega \propto | {\bf \vec{k}} |^z$, and for a 1D conductor
one knows that $z=1$. Hence, 
$${\rm Im}\; G ( \omega , k_F )
\propto \omega^{-z + {\alpha}}
\sim \omega^{-1 + {\alpha}}.$$
The flow of the hopping terms under a Renormalization Group
scaling of the cutoff is \cite{1D}:
\begin{equation}
\frac{\partial ( {\bf t} / \Lambda )}{\partial l} =
\left\{ \begin{array}{lr} - {\alpha} &{\rm localized \, \, \, hopping} \\
1 - {\alpha} &{\rm extended \, \, \, hopping} \end{array} \right.
\label{scaling}
\end{equation}
where ${\bf t}$ denotes a hopping term, between localized or extended states.
In the latter case,
the hopping becomes
an irrelevant variable\cite{W90} for ${\alpha} > 1$.

\section{Layered models}
\subsection{Vanishing density of states at the Fermi level.} 
The simplest two dimensional model for interacting electrons where
it can be rigourously shown that the couplings acquire logarithmic
corrections in perturbation theory is a system of Dirac fermions
($\epsilon_k = v_F | {\bf \vec{k} |}$), with Coulomb, $1 / |
{\bf \vec{r} - \vec{r}'} |$, interaction. This model can be used
to describe isolated graphene planes\cite{GGV94,GGV99}, and can 
help to understand the anomalous behavior of graphite observed
in recent experiments\cite{Yetal96,STK01}. Nodal quasiparticles
in d-wave superconductors are also describable in these terms.

In order to apply the procedure outlined in the previous section, one 
needs the Fourier transform of the interaction, 
$$V_{eff}({\bf \vec{k}})=e^2 / ( \epsilon_0
| {\bf \vec{k}} |),$$
where $e$ is the electronic charge, and 
$\epsilon_0$ is the dielectric constant, and the susceptibility
of the electron gas. For a single graphene plane, this quantity
has been computed in \cite{GGV94} and is:
\begin{equation}
\chi_0 ({\bf \vec{k}}, \omega ) = \frac{1}{8} \frac{ | {\bf
\vec{k}}|^2} {\sqrt{v_F^2 |{\bf \vec{k}}|^2 - \omega^2 }}
\label{chigraphite}
\end{equation}
These expressions need to be inserted 
in equations (\ref{veff}) and (\ref{propagator}).
%Alternatively, we can use the Random Phase Approximation (RPA), 
%and include the effects
%of interplane screening:
%
%\begin{equation} 
%\begin{array}{ll}
%\chi_{RPA} ({\vec{\bf k}}, \omega  ) &= \\
%%\frac{2 \pi e^2}{ \epsilon_0 | {\bf k} | }
%& \frac{\sinh ( | {\bf \vec{k}} | d )}
%{\sqrt{ \left[ \cosh ( | {\vec{\bf k}} | d ) + \frac{2 \pi e^2}
%{ \epsilon_0 | {\bf k} |}
%\sinh ( | {\vec{\bf k}} | d )
%\chi_0 ( {\vec{\bf k}} , \omega ) \right]^2 - 1 }} \end{array} 
%\label{RPA}
%\end{equation}
%where $d$ is the interplane spacing. 

For simplicity, we consider the expression in Eq.(\ref{chigraphite}),
as it allows us to obtain analytical results. 
The imaginary part, ${\rm Im} \chi_0
( {\bf \vec{k}} , \omega )$, is different from zero if $\omega >
v_F | {\bf \vec{k}} |$. We cut off the spatial
integrals at a scale, $L$, of the order of the electronic
wavefunctions involved in the tunneling. Performing the same computation as
in the case of D=1 we obtain an expression
similar to that in Eq.(\ref{1D}) except that the prefactor
$( U / E_F )^2$ is replaced by the squared of  effective coupling constant
of the model, $e^4 / ( \epsilon_0 v_F )^2$.
Thus, also  in the graphene model, the propagators acquire an anomalous 
dimension what was advocated  in \cite{GGV94} as pointing out to a departure
of the model from Fermi liquid behavior. As in 1D,
the value of the exponent $z$ which relates length and time scales
is $z=1$. The scaling of the hoppings now are:
\begin{equation}
\Lambda \frac{\partial ( {\bf t} / \Lambda )}{\partial \Lambda} =
\left\{ \begin{array}{lr} - 
1 - {\alpha} &{\rm localized \, \, \, hopping} \\
1 - {\alpha} &{\rm extended \, \, \, hopping} \end{array} \right.
\label{scaling_graphite}
\end{equation}
The extra constant in the first equation with respect to Eq. (\ref{scaling})
reflects the vanishing of the density of states at the Fermi level for
two dimensional electrons with a Dirac dispersion relation.

In graphite, the dimensionless coupling constant,  $e^2 / v_F$,
is of order unity. Under renormalization, it flows towards
zero\cite{GGV94}.
Thus, despite the departure from Fermi liquid behavior,
interplane tunneling is a relevant variable
and a coherent out of plane transport in clean graphene samples
should be observed.
The former picture can change in the presence of disorder that
is known to change the anomalous dimension of the fields\cite{GGV01}.

\subsection{Saddle point in the density of states.}
%{\it Saddle point in the density of states.}
The Fermi surface of most hole-doped cuprates is close to a saddle point
of the dispersion relation\cite{vhexp}.  
The possible relevance of this fact to the
superconducting transition as well as to the anomalous behavior of the
normal state was put forward in the early times of the cuprates 
and gave rise to the so called Van Hove scenario\cite{vhscenario}.

We shall apply the  mechanism described in section  II
to study the interlayer hopping of two electronic systems described by the
{\bf t--t'} Hubbard model and whose Fermi surface lies close to a Van Hove 
singularity.

The {\bf t-t'}-Hubbard
model has the dispersion relation
\begin{eqnarray}
\varepsilon({\vec{\bf k}})& =& -2{\bf t}\;[ \cos(k_x a)+\cos(k_y a)] \nonumber \\
&-&2{\bf t'}\cos(k_x a)\cos(k_y a)
\;\;\;, 
\label{disp0}
\end{eqnarray}
This dispersion relation has two inequivalent saddle points at 
A $(\pi,0)$ and  B $(0,\pi)$.
The Van Hove model in its simplest 
formulation is obtained by 
assuming that for fillings such that
the Fermi line lies close to the singularities, the majority of
states participating in the interactions will come from regions in the
vicinity of the saddle points. In the present context the model is
further reinforced by the fact that
the momentum dependence of the out of plane hopping in the 
cuprates 
tends to suppress tunneling from points away from the singularities.  
Taylor expanding Eq. (\ref{disp0}) around
the two points gives the effective relation
\begin{equation}
\varepsilon_{A,B} ({\vec {\bf k}} ) \approx \mp ( {\bf t} \mp 2 {\bf t'} ) k_x^2 
\pm ( {\bf t} \pm 2 {\bf t'} ) k_y^2  \;\;\;,
\label{disp}
\end{equation}
The dynamical exponent (\ref{scaling_g0}) in this case is z=2.
The dispersion relation (\ref{disp})
allows to formulate a renormalizable effective model
based on the hamiltonian:
\begin{eqnarray}
{\cal H} &= &\sum_{i=A,B;k,s}^{\epsilon_{i,k} < \Lambda_0}
 \epsilon_{i,k} c^\dag_{k,i,s}
c_{i,k,s} \nonumber \\ &+ &\sum
u_{i,i';s,s'} c^\dag_{i,k,s} c_{i',k',s'}
c^{\dag}_{i'',k'',s''} c_{i''',k''',s''' }\;,
\end{eqnarray}
%\begin{equation}
%{\cal H} = \sum_{i=A,B;k,s}^{\epsilon_{i,k} < \Lambda_0}
% \epsilon_{i,k} c^\dag_{k,i,s}
%c_{i,k,s} + \sum_{IJ}U_{IJ}n_In_J \; ,
%\end{equation}
where $\Lambda_0$ is a high energy cutoff which sets the limit of 
validity of the effective description.
%$n_I$ is the density
%operator, and the interaction $U_{IJ}$
%encodes the various low--energy four fermi interactions
%that the model can support\cite{AGGV}.  
The particle--hole susceptibility has been  computed in \cite{GGV96}:
%
%\begin{equation}
%{\rm Re} \: \chi ({\bf{\bf k}}, \omega) = 
%\frac{c}{2 \pi^2}
%\frac{U^2}{t} \;\left[ \log \: \frac{\left| \varepsilon ({\bf k})
%\right| \Lambda}{\left| \omega^2 - \varepsilon ({\bf k})^2
% \right|}  +   \frac{\omega}
% {\varepsilon ({\bf k}) } \; \log \: \left| \frac{\omega -
% \varepsilon ({\bf k}) }{\omega + \varepsilon ({\bf k}) }
% \right|  \right]
%\label{re}
%\end{equation}
%
\begin{equation}
{\rm Im} \: \chi ({\vec{\bf k}}, \omega) = 
\frac{1}{4 \pi\varepsilon ({\vec{\bf k}})} \left(\left| \omega+\varepsilon 
({\vec{\bf k}})
\right| - \left| \omega-\varepsilon ({\vec{\bf k}}) \right|\right)\;,
\end{equation}
where $\varepsilon({\vec{\bf k}})$ is the dispersion relation (\ref{disp}).
%$\Lambda$ is an energy cutoff, and
%$c \equiv 1/\sqrt{1 - 4(t'/t)^2} \;.$

The long time dependence of the Green's function is determined
by the low energy behavior of $\chi$: 
$$\lim_{\omega \to 0}{\rm Im} \: \chi ({\vec{\bf k}}, \omega) \sim 
\frac{\omega}{\varepsilon ({\vec{\bf k}})}$$
Inserting this expression in eqs.(\ref{propagator}) and
(\ref{veff}), 
we can see that, irrespective of the details
of the interaction,  in the presence of
a Van Hove singularity the exponent ${\alpha}$ in the
time dependence of the Green's function goes as:
$$\lim_{\Omega \rightarrow
\infty} {\alpha} \propto \log ( L )\;\;,$$
where $L$, as before, is the length scale which characterizes
the wavefunction of the tunneling electron, and $\Omega \propto
L^2$ is the size of the integration region in Eq.(\ref{propagator}).
The details of the anomalous dimension of the propagator in this case
depend on the nature of the interactions which determine
$V_{eff}({\bf \vec{k}},\omega)=V^2({\bf \vec{k}})
{\rm Im} \chi ( {\bf\vec{k}} , \omega )\;$.
%$$
%\big<\delta V_\omega^2\big>
%=\int\frac{d^2{\bf k}}{(2\pi)^2} V_eff({\bf k},\omega)\\ \nonumber
%\sim {\alpha} \omega\;.$$

To make contact with  the work  of Turlakov and Leggett \cite{TL01} 
about the $c$-axis interlayer tunneling in the cuprates,
we have computed the parameter ${\alpha}$ for 
different possible interactions. In \cite{TL01} several models, describing
the voltage noise, are examined to estimate the parameter ${\alpha}$ in order
to describe the $c$-axis transport behavior. Here we consider three examples of potential:
a long-range Coulomb interaction, a short-range Coulomb interaction and the intermediate
situation between them, the Thomas-Fermi screened potential. The parameter ${\alpha}$
is calculated in the three cases.

a) Unscreened Coulomb potential,
$ V({\bf \vec{k}})=\frac{2\pi e^2}{\epsilon_0 \vert 
{\bf \vec{k}}\vert}$. 
Due to the highly singular interaction, the $\omega$
dependence of the effective potential is not linear and ${\alpha}$
is not well defined in this case. We can still analyze the
tunneling by computing 
the effective potential Eq. (\ref{veff}). It  is
computed to be:
$$V_{eff}({\bf \vec{k}},\omega)=\frac{ e^4}{\epsilon_0^2 {\bf t}}
\left[ 1+\log\left(\frac{\omega}{\omega_0}\right)
\right]$$
where $\omega_0 = | E_F - E_{VH} |$ is a low-energy 
cutoff, $E_F$ is the Fermi energy and the position of the saddle point is
at $E_{VH}$. $\omega_0$ keeps the chemical potential away from the
singularity and is required to avoid infrared divergences in the 
integrals\cite{GGV96}. From Eq. (\ref{propagator}) it can be seen that 
the tunneling in this case is exponentially suppressed.

b) Consider now a
screened interaction of the  Hubbard type, $V({\bf \vec{k}})=U a^2$,
where $a$ is the lattice unit. In this case the effective 
potential has a linear term in $\omega$ and the formalism
described proceeds strightforward to give a value of the
exponent $\alpha$ which is
$${\alpha}=\frac{4U^2}{\pi {\bf t}^2}\frac{1}{(2\pi)^2}K_M
\left[1+\frac{1}{2}
\log\left(\frac{\Lambda}{\omega}\right)\right]\;\;,
$$
where $\Lambda$ is a high-energy cutoff, and 
$$K_M=\frac{1}{2}\log \left\vert\frac{k_x+k_y}{k_x-k_y}\right\vert\;,$$
where $k_x , k_y \sim \sqrt{\omega_0}$. The RG analysis of
\cite{GGV96,GGV97} shows that the coupling constant
$U$ of the model renormalizes to large values
making the interlayer tunneling irrelevant.

c) Finally we will analyze the case of a
Thomas--Fermi screened potential which is an intermediate
situation with respect to the previous cases. 
$$ V({\bf \vec{k}})^2=\frac{(2\pi e^2)^2}{\epsilon_0^2(k^2 + k_{TF}^2)}\;.$$
In this case $\alpha$ can also be defined and it
is computed to be
$$\alpha\sim\frac{2e^4 k_{TF}^2}
{\epsilon_0\pi {\bf t}^2}\log^2(k_{TF} / k)\;,$$
where $k_{TF}$ is the Thomas-Fermi wavevector,
and, $k$ is a momentum cutoff  $k \sim \sqrt{\omega_0}$. 
In this case the longer range of the  interaction makes the
divergence softer as the dependence of the cutoff is as a
squared log instead of a linear log.

In all the three  cases studied ${\alpha}$ diverges as $E_F 
\rightarrow E_{VH}$. Thus, interlayer hopping is an 
irrelevant variable, and scales towards zero as the
temperature or frequency is decreased. The additional
logarithmic dependence found can be seen as a manifestation
of the $\log^2$ divergences
which arise in the treatment of this model\cite{GGV96}.
Note that, as in the graphene case, the coupling
constants are also energy dependent, but in the Van Hove
case they have an unstable flow and  grow at
low energies, suppressing even further the interlayer
tunneling.
This behavior of the interlayer hopping is in agreement with
that obtained in Ref. \cite{TL01} where it is found that $c$-axis interlayer
tunneling is suppressed by voltage fluctuations.

In the case of a layered system,  
we can use the Random Phase Approximation (RPA), 
and include the effects
of interplane screening\cite{quinn}:

\begin{equation} 
\begin{array}{ll}
\chi_{RPA} ({\vec{\bf k}}, \omega  ) &= \\
%\frac{2 \pi e^2}{ \epsilon_0 | {\bf k} | }
& \frac{\sinh ( | {\bf \vec{k}} | d )}
{\sqrt{ \left[ \cosh ( | {\vec{\bf k}} | d ) + \frac{2 \pi e^2}
{ \epsilon_0 | {\bf k} |}
\sinh ( | {\vec{\bf k}} | d )
\chi_0 ( {\vec{\bf k}} , \omega ) \right]^2 - 1 }} \end{array} 
\label{RPA}
\end{equation}
where $d$ is the interplane spacing. 

A numerical computation of \ref{RPA}
provides the effective potential 
Eq. (\ref{veff}). The imaginary part $ \chi_{RPA}$
gives the quantity usually known as the loss function
that can be experimentally  determined.
This calculation is shown in Fig. [\ref{loss}].

\begin{figure}
\begin{center}
\resizebox{6cm}{!}{\includegraphics[width=7cm]{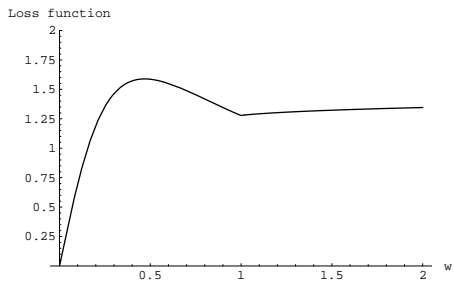}} 
%\mbox{\psfig{file=caxisfig1.eps,width=.35\textwidth}}
\end{center}
\caption{Effective potential calculated as a function
of the energy for fixed ${\bf \vec{k}}$ for a system of Van Hove
layers coupled by Coulomb interaction .} 
\label{loss}
\end{figure}

The calculated effective potential  compares well with the experimental plots
of the  energy loss function of Bi$_2$Sr$_2$CaCuO$_2$, measured
by transmission energy loss spectroscopy in the low energy region \cite{curvaexp},  
what reveals that the Van Hove model is also compatible with transport experiments.
The results for the parameter
${\alpha}$ extracted from the numerical computation
are in qualitative agreement with the analytical
expressions given in cases of  b) screened Hubbard interaction  and 
c) Thomas-Fermi screened potential, studied above.

In a recent publication\cite{TL03} it is argued that Umklapp
processes are crucial to determine the spectrum of density
fluctuations, especially in two dimensional systems. It is
worth to notice that the dispersion relation 
of the Van Hove model (\ref{disp}) is a case
where Umklapp processes enhance significantly the response
function at low frequencies and wavevectors.

\section{Conclusions.}

We have discussed the supression of interlayer tunneling by
inelastic processes in two dimensional systems in the clean limit.
Our results suggest that, when perturbation theory
for the in--plane interactions leads to logarithmic
divergences,
the out of plane tunneling acquires
a non trivial energy dependence. This anomalous scaling
of the interlayer hopping can make it irrelevant, at
low energies, if the in--plane interactions are sufficiently
strong. A well known problem where this non trivial scaling takes place 
is the tunneling between
one dimensional Luttinger liquids\cite{W90,KF92,1D}

In two dimensions, the scaling towards zero of
the out of plane hopping  is always the case if the Fermi level
of the interacting electrons lies at a van Hove 
singularity (note that the Fermi level can, in certain
circumstances, be pinned to the singularity\cite{GGV96}). 
In this situation, insulating behavior in the
out of plane direction is not incompatible with
gapless
 or even superconducting 
in--plane properties. If the Fermi level is not tuned
to the singularity, the scaling presented here is only valid for
energies larger than the distance of the Fermi level to the
Van Hove singluarity. Within this range of energies or
temperatures, insulating behavior can be expected,
and Fermi liquid (coherent) behavior will set in at lower energies. The
appearance of instabilities can render this low energy regime unobservable.

Clean graphene planes show the opposite behavior, as electron-electron
interactions become irrelevant at low energies.

%\narrowtext
%\section{Acknowledgements}
{\it Acknowledgements}.
The financial support of the MCYT (Spain), through
grant no. MAT2002-04095-C02-01
is gratefully acknowledged.

%\begin{thebibliography}{}

%\end{thebibliography}
\end{document}